\documentclass[aps,twocolumn,pra,superscriptaddress,amsmath,amssymb]{revtex4-1}

\ifx\pdfoutput\undefined
\usepackage[dvipdfmx]{graphicx}
\usepackage[dvipdfmx]{hyperref}
\usepackage[dvipdfmx]{color}
\else
\usepackage{color}
\usepackage{graphicx}
\usepackage{hyperref}
\fi
\newcommand{\ket}[1]{\left|#1\right>}

\newcommand{\means}[1]{\langle#1\rangle}

\usepackage{soul}
\soulregister\cite7
\soulregister\ref7
\soulregister\pageref7

\DeclareRobustCommand{\nchange}[2]{\ifmmode{{\textrm{\setul{}{1pt}\setstcolor{blue}\st{$\displaystyle#1$}}}}\else{{\setul{}{1pt}\setstcolor{blue}\st{#1}}}\fi\ \textcolor{blue}{#2}}
\begin{document}

\title{
  Low temperature properties in the Bilayer Kitaev model
}

\author{Hiroyuki Tomishige}
\affiliation{
  Department of Physics, Tokyo Institute of Technology,
  Meguro, Tokyo 152-8551, Japan
}
\author{Joji Nasu}
\affiliation{
  Department of Physics, Tokyo Institute of Technology,
  Meguro, Tokyo 152-8551, Japan
}
\affiliation{
  Department of Physics, Yokohama National University,
  79-5 Tokiwadai, Hodogaya, Yokohama 240-8501, Japan
}

\author{Akihisa Koga}
\affiliation{
  Department of Physics, Tokyo Institute of Technology,
  Meguro, Tokyo 152-8551, Japan
}

\begin{abstract}
  The ground state of the bilayer Kitaev model with the Heisenberg-type interlayer exchange interaction
  is investigated by means of the exact diagonalization.
  Calculating the ground-state energy, local quantity defined on each plaquette, and
  dynamical spin structure factor, we obtain results suggesting the existence of a quantum phase transition
  between the Kitaev quantum spin liquid (QSL) and dimer singlet states
  when the interlayer coupling is antiferromagnetic.
  On the other hand, increasing the ferromagnetic interlayer coupling, there exists no singularity
  in the physical quantities, suggesting that
  the $S=1/2$ Kitaev QSL state realized in each layer adiabatically
  connects to another QSL state realized in the $S=1$ Kitaev model.
  Thermodynamic properties are also studied by means of the thermal pure quantum state method.
\end{abstract}
\maketitle

%%%%%%%%%%%%%%%%%%%%%%%%%%%%%%%%%%%%
\section{Introduction}
%%%%%%%%%%%%%%%%%%%%%%%%%%%%%%%%%%%%

Recently, quantum spin liquid (QSL),
which is an exotic magnetic state with no long-range order
even at zero temperature,
has attracted much interest in modern condensed matter physics~\cite{Anderson1973}.
One of the celebrated examples is a ground state of the one-dimensional
quantum spin Heisenberg model,
where the gap formation reflects the topological nature of spins~\cite{PhysRevLett.50.1153,Haldane};
low-energy excitations are gapless for the half-integer spins
and gapped for the integer spins.
These interesting ground-state properties have been examined numerically~\cite{Nightingale,White}.
Furthermore, its qualitative difference has been discussed by studying
the interpolating models such as
the double spin $S=1/2$ chains coupled by the ferromagnetic exchange interaction~\cite{Hida1991,Watanabe1993}.
In the system, the introduction of the interchain interaction immediately induces the gapped state.
This implies that the gapless QSL state possesses an instability against the perturbations,
while the gapped state is more stable.

Another interesting example is the QSL state in the two-dimensional Kitaev model
with three kinds of bond-dependent Ising interactions for $S=1/2$ spins~\cite{Kitaev2006}.
It is known that this model is exactly solvable and
spin degrees of freedom are decoupled to itinerant Majorana fermions and $Z_2$ fluxes.
Although elementary excitations are gapless, there exists a gap in the spin excitations with short-range spin correlations~\cite{Baskaran2007,Knolle2014}.
Therefore, this gapless QSL state should be stable, in contrast to the QSL state in one dimension.
So far, the stability of the QSL in the $S=1/2$ Kitaev model has been examined
in the presence of perturbations
such as magnetic field~\cite{PhysRevB.83.245104,yadav_kitaev_2016,Nasu2018},
Heisenberg interactions~\cite{Chaloupka2010,Chaloupka2013,Katukuri2014,Yamaji2014,Yamaji2016},
inverse of the spin-orbit coupling~\cite{Nakauchi} and
interlayer couplings~\cite{tomishige2018,pKoga2018,seifert2018bilayer}.

In addition to the $S=1/2$ Kitaev model, its generalizations
to arbitrary spin amplitudes $S$ have also short-ranged spin correlations,
%\red{[also have no long-ranged spin correlations?]},
suggesting gapped spin excitations~\cite{PhysRevB.78.115116}.
However, less is known for the generalized spin-$S$ Kitaev models, particularly for their spin excitation spectra.
In our previous paper~\cite{koga2018ground},
the $S=1$ Kitaev model has been considered and
it has been clarified the presence of low energy excitations from the ground state
and two energy scales observed separately in the thermodynamic quantities.
These results naively expect that the generalized Kitaev model has common magnetic properties.
On the other hand, it has been reported that, in the anisotropic Kitaev model
with one of the three kinds of bonds being large,
the effective Hamiltonian depends on the parity of $2S$;
its ground state is quantum (classical) for half-integer (integer) spins~\cite{Minakawa}.
Therefore, it is instructive to clarify
the connection between these different spin sectors in the QSL states of the isotropic Kitaev model.

In our paper, we consider the bilayer Kitaev model with the Heisenberg-type interlayer exchange interactions~\cite{tomishige2018}.
When the interlayer ferromagnetic exchange interactions are large enough,
the system is reduced to the $S=1$ Kitaev model~\cite{koga2018ground}.
Therefore, this bilayer model has an advantage in discussing how
the QSL state for the $S=1/2$ Kitaev model is connected
to that for the $S=1$ Kitaev model.
By using numerical techniques, we study the bilayer Kitaev model
to discuss the essence of the ground state properties
in the generalized Kitaev model~\cite{PhysRevB.78.115116,koga2018ground,Oitmaa,Minakawa}.

The paper is organized as follows.
In Sec.~\ref{sec:2},
we introduce the bilayer Kitaev model and derive
the effective model in the strong ferromagnetic interlayer coupling limit.
In Sec.~\ref{sec:3},
we study the ground-state properties by means of the exact diagonalization.
Calculating thermodynamic quantities by means of
the thermal pure quantum (TPQ) state method~\cite{Sugiura2012,Sugiura2013},
we discuss a double peak structure in the specific heat characteristic of the Kitaev model.
A summary is given in the final section.

%%%%%%%%%%%%%%%%%%%%%%%%%%%%%%%%%%%%
\section{Model and methods}\label{sec:2}
\begin{figure}[htb]
\centering
\includegraphics[width=\columnwidth]{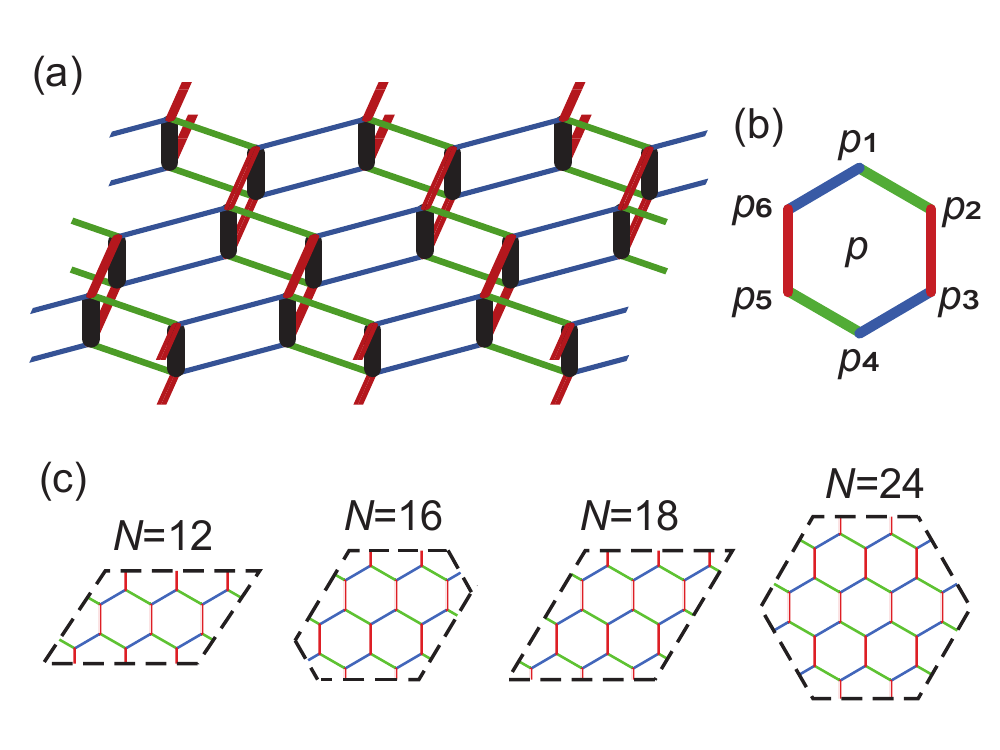}
\caption{
 (a) Bilayer Kitaev model on the honeycomb lattice.
 Red, blue, and green lines represent $x$-, $y$-, and $z$-bonds, and
 black lines interlayer bonds with the Heisenberg interactions.
 (b) Top view of the bilayer Kitaev model with the definition of
	index of plaquette, $p$, associated with dimer sites, $p_1$--$p_6$.
 (c) Finite size clusters used in the ED calculations.
}
\label{fig:model}
\end{figure}

%%%%%%%%%%%%%%%%%%%%%%%%%%%%%%%%%%%%
We consider here the bilayer Kitaev model~\cite{tomishige2018},
where two monolayer Kitaev models are coupled by the Heisenberg interactions.
The Hamiltonian is described as,
\begin{align}
 {\cal H} = -  J_K\sum_{\means{ij}_\alpha,n} S_{i,n}^\alpha S_{j,n}^\alpha + J_H \sum_{i} {\bf S}_{i,1}\cdot{\bf S}_{i,2},\label{eq:1}
\end{align}
where $S_{i,n}^\alpha$ is the $\alpha(=x,y,z)$ component of
the $S=1/2$ operator at the $i$th site on the $n(=1,2)$th layer.
The inplane bond-dependent Ising-type interaction $J_K (>0)$ is defined on the three kinds of bonds such as
$x$-, $y$-, and $z$-bonds linking nearest neighbor sites.
$J_H$ is the Heisenberg interlayer coupling.
The model is schematically shown in Fig.~\ref{fig:model}(a).
%%%%%%%%%%%%%%%%%%%%%%%%%%%%%%%%%%%%
%%%%%%%%%%%%%%%%%%%%%%%%%%%%%%%%%%%%
In the paper, we mainly study ground state properties
in the bilayer Kitaev model with ferromagnetic interlayer Heisenberg interactions $J_H<0$,
and some results with antiferromagnetic interactions $J_H>0$ are shown for comparison.

What is the most important is that
this bilayer Kitaev model is reduced to the interesting models in certain limits.
When $J_H=0$, two layers are decoupled and the system is represented by
two exactly solvable monolayer Kitaev models.
It is known that, in the monolayer model,
a QSL state is realized at zero temperature, where
the fractionalization of quantum spins yields
gapless Majorana excitations and gapful spin excitations.
%In addition, it is known that the difference of the energy scales induces double peak structure
%in the thermodynamic quantities~\cite{Nasu2014,Nasu2015}.
%On the other hand, $J_H/J_K\rightarrow +\infty$,
%each spin is coupled only by the antiferromagnetic interlayer coupling
%and the direct product of dimer singlets is realized.
%The competition between these two states has been examined and
%the first order phase transition occurs around $J_H/J_K\sim 0.05$.
On the other hand, in the case $J_H\rightarrow -\infty$,
the spin triplet state is realized at each site.
In the case, the effective Hamiltonian is shown to be described by
the $S=1$ Kitaev model as
%%%%%%%%%%%%%%%%%%%%%%%%%%%%%%%%%%%%%%%%%%%%%%%%%%%
\begin{align}
  {\cal H}^{\mathrm{ eff}} = -  J_{\rm eff}\sum_{\means{ij}_\alpha} \tilde{S}_i^\alpha \tilde{S}_j^\alpha,
  %  +\frac{1}{4}J_HN,
  \label{eq:eff}
\end{align}
%%%%%%%%%%%%%%%%%%%%%%%%%%%%%%%%%%%%%%%%%%%%%%%%%%%
where $J_{\rm eff}=J_K/2$ and $\tilde{S}_i^\alpha$ is the $\alpha$ component of the $S=1$ operator at the $i$th site.
It is clarified that
its ground state is nonmagnetic~\cite{PhysRevB.78.115116}, and
its excitation is suggested to be gapless~\cite{koga2018ground}.
Furthermore, it has been clarified that
the double peak structure appears in the specific heat~\cite{koga2018ground,Oitmaa}.
In the paper, we systematically examine the bilayer Kitaev model to clarify
how the nonmagnetic states for the $S=1/2$ and $S=1$
Kitaev models are connected to each other.

We briefly comment on the conserved quantities in the bilayer Kitaev model~\cite{tomishige2018}.
We consider the following operator defined on a certain plaquette $p$, as
\begin{eqnarray}
  X_p&=&W_{p1}W_{p2},\\
  W_{pn}&=&\sigma_{p_1,n}^x \sigma_{p_2,n}^y \sigma_{p_3,n}^z
  \sigma_{p_4,n}^x \sigma_{p_5,n}^y \sigma_{p_6,n}^z,
\end{eqnarray}
where $\sigma^\alpha_{p_i,n}$ is the $\alpha$ component of the Pauli matrix
at the $i(=1,2,3,4,5,6)$th site on the plaquette $p$ in the $n$th layer [see Fig.~\ref{fig:model}(b)].
We note that $W_{pn}$ defined on the $n$th layer is a local conserved quantity
for the monolayer Kitaev model~\cite{Kitaev2006}.
The operator $X_{p}$ for each plaquette $p$
commutes with the bilayer Hamiltonian Eq.~(\ref{eq:1}),
which guarantees no long range magnetic order in the system~\cite{tomishige2018}.
Given that $X_p^2=1$, $X_p$ is confirmed to be a $Z_2$ conserved quantity.
In addition, there exists a global parity symmetry for the number of local singlets and triplets
on the sites in the bilayer system~\cite{tomishige2018}.
The corresponding parity operators are
\begin{eqnarray}
  P_S&=&\exp\left[i\pi\sum_i|s_i\rangle\langle s_i|\right],\\
  P_{T_\alpha}&=&\exp\left[i\pi\sum_i|t^\alpha_i\rangle\langle t^\alpha_i|\right],
\end{eqnarray}
where $|s_i\rangle$ and $|t^\alpha_i\rangle$ are
singlet and $\alpha(=x,y,z)$ triplet states on the $i$th dimer site.
Since the operators ${\cal H}, X_p, P_S$, and $P_{T_\alpha}$ commute with each other,
the subspace of the Hamiltonian should be specified by
${\cal S}[\{p_{t_\alpha}\}, \{x_p\}]$,
where $p_{t_\alpha}(=\pm 1)$ and $x_p(=\pm 1)$ are the eigenvalues of
$P_{T_\alpha}$ and $X_p$, respectively.
Note that the eigenvalue of $P_S$ is uniquely
determined by $\{p_{t_\alpha}\}$ and the number of dimer sites, $N$.

%\subsection{Numerical methods}\label{methods}
%Making use of the above conserved quantities,
%we reduce the dimension of the Hamiltonian matrix and
The original Hamiltonian is explicitly represented by the set of smaller matrices defined
in the subspace ${\cal S}[\{p_{t_\alpha}\}, \{x_p\}]$, which enables us to
perform the exact diagonalization (ED) with the finite size clusters up to 24 sites.
When we consider several clusters shown in Fig.~\ref{fig:model}(c),
the ground state written by $\ket{\Phi_0}$ always belongs to the subspace
${\cal S}[\{p_t^\alpha=1\}, \{x_p=1\}]$
and is non-degenerate when $J_H\neq 0$.
We then evaluate the static quantities $E_0(=\langle {\cal H}\rangle)$ and $\langle W_{pn}\rangle$
to study quantum phase transitions in the system.
To discuss the spin excitation in the system,
we also calculate the dynamical spin structure factor,
which is defined as
%%%%%%%%%%%%%%%%%%%%%%%%%%%%%%%%%%%%%%
\begin{eqnarray}
S^{\alpha\beta}_n({\bf q},\omega)
&=&-\frac{1}{\pi}{\rm Im} \langle\Phi_0\rvert S^{\alpha}_{-{\bf q}n} \frac{1}{\omega_++E_0-\cal H}
S^\beta_{{\bf q}n}\vert\Phi_0\rangle,\\
 S^\alpha_{{\bf q}n}&=&\frac{1}{\sqrt{N}}\sum_i S^\alpha_{in}e^{i{\bf q}\cdot {\bf r}_i},
\end{eqnarray}
where $\omega_+=\omega+i\delta^+$.
%%%%%%%%%%%%%%%%%%%%%%%%%%%%%%%%%%%%%%
The quantity can be evaluated by the continued fraction
expansion~\cite{dagotto1994correlated,gagliano1987dynamical} as,
%%%%%%%%%%%%%%%%%%%%%%%%%%%%%%%%%%%%%%
\begin{align}
S^{\alpha\beta}_n({\bf q},\omega)
= \frac{\langle\Phi_0\rvert S^{\alpha}_{-{\bf q}n} S^\beta_{{\bf q}n} \vert\Phi_0\rangle}{\displaystyle \omega_+-a_1-\frac{b_2^2}{\displaystyle \omega_+-a_2-\frac{b_3^2}{\omega_+-\cdots}}},
\end{align}
%%%%%%%%%%%%%%%%%%%%%%%%%%%%%%%%%%%%%%
where $a_i$ and $b_i$ are the $i$th diagonal and off-diagonal elements of
the Hamiltonian tridiagonalized by the Lanczos method
with the initial vector $S^\alpha_{{\bf q}n}\ket{\Phi_0}$.
%/\sqrt{\bra{\phi_0}S^{\nu}_{-qn}S^\nu_{qn}\ket{\phi_0}}$.
Here, $S^{\alpha\beta}_n({\bf q},\omega)\ (\alpha\neq\beta)$ are zero
due to the existence of the local conserved quantity in the system.

We also study finite temperature properties
to clarify how unique features in the generalized Kitaev models,
namely, double peak structure in the specific heat
and the plateau in the entropy at $\ln (2S+1)/2$,
appear in the bilayer Kitaev model.
To this end, we make use of the TPQ
state method~\cite{Sugiura2012,Sugiura2013},
which is one of the powerful methods to evaluate thermodynamic quantities in the system.
In the calculations, we prepare more than 10 random vectors
for the initial states, and thermodynamic quantities are deduced by
averaging the values generated by these states.

%%%%%%%%%%%%%%%%%%%%%%%%%%%%%%%%%%%
\section{Results}\label{sec:3}
%%%%%%%%%%%%%%%%%%%%%%%%%%%%%%%%%%%

%%%%%%%%%%%%%%%%%%%%%%%%%%%%%%%%%%%%
%\subsection{Ground state properties}
%%%%%%%%%%%%%%%%%%%%%%%%%%%%%%%%%%%%
First, we start with the ground state properties of the bilayer Kitaev model.
%%%%%%%%%%%%%%%%%%%%%%%%%%%%%%%%%%%%
\begin{figure}[htb]
\centering
\includegraphics[width=\columnwidth]{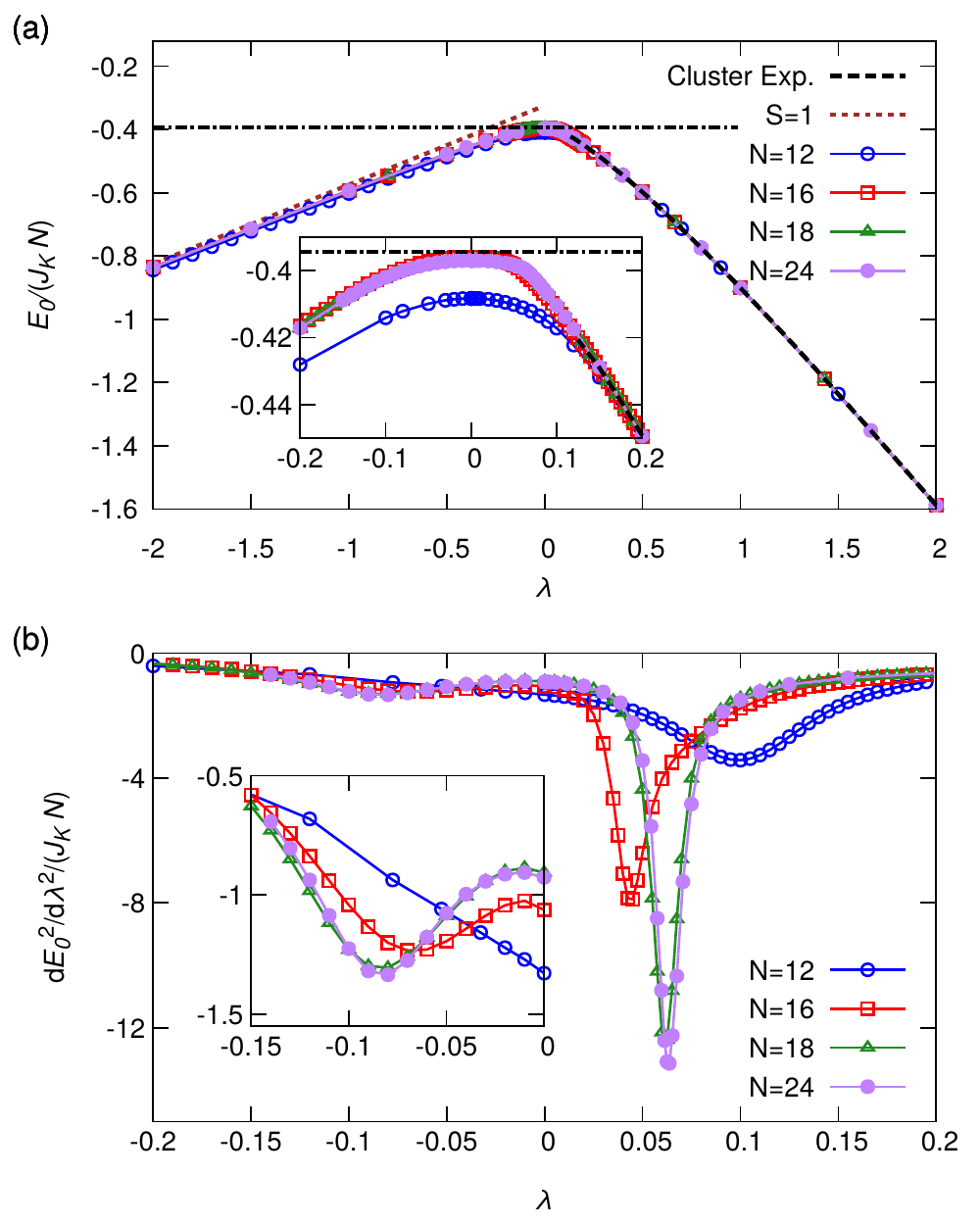}
\caption{
(a) Ground energy per dimer as a function of $\lambda$
in the bilayer Kitaev model.
The dashed-dotted line represents analytically calculated
ground energy of the Kitaev model.
(b) Second-order differential of the ground energy per dimer.
}
\label{fig:emin}
\end{figure}
%%%%%%%%%%%%%%%%%%%%%%%%%%%%%%%%%%%%
Figure~\ref{fig:emin}(a) shows the ground energy $E_0$
as a function of $\lambda ( = J_H/J_K)$
obtained by the exact diagonalizations for several clusters.
When $\lambda>0.2$,
the present results are in a good agreement with the results
obtained by the dimer expansion~\cite{tomishige2018}, meaning that
the dimer singlet state is realized in the region.
On the other hand, increasing the ferromagnetic interlayer coupling,
the ground state energy approaches that for the $S=1$ Kitaev model
$E/N=J_H/4-0.65 J_{\rm eff}$, as shown in Fig.~\ref{fig:emin}(a).
We note that the system size dependence appears around $\lambda=0$,
which is clearly shown in the inset of Fig.~\ref{fig:emin}(a).
Since ground state properties for a finite cluster are sensitive to the low lying excitations,
this finite size effect may suggest that the gapless QSL state is stable even for finite $\lambda$.
To examine this,
we also show the second-order derivative of $E_0$ in Fig.~\ref{fig:emin}(b).
In the positive $\lambda$ region, we find that the peak structure
around $\lambda\sim 0.05$ develops with increasing $N$.
This strongly suggests the existence of the quantum phase transition.
On the other hand, in the negative $\lambda$ region,
we could not find such a clear signature in the curves although
a broad and small peak is found around $\lambda\sim -0.1$.
Therefore, the results suggest the absence of 
quantum phase transition in the negative $\lambda$ case.

\begin{figure}[htb]
\centering
\includegraphics[width=\columnwidth]{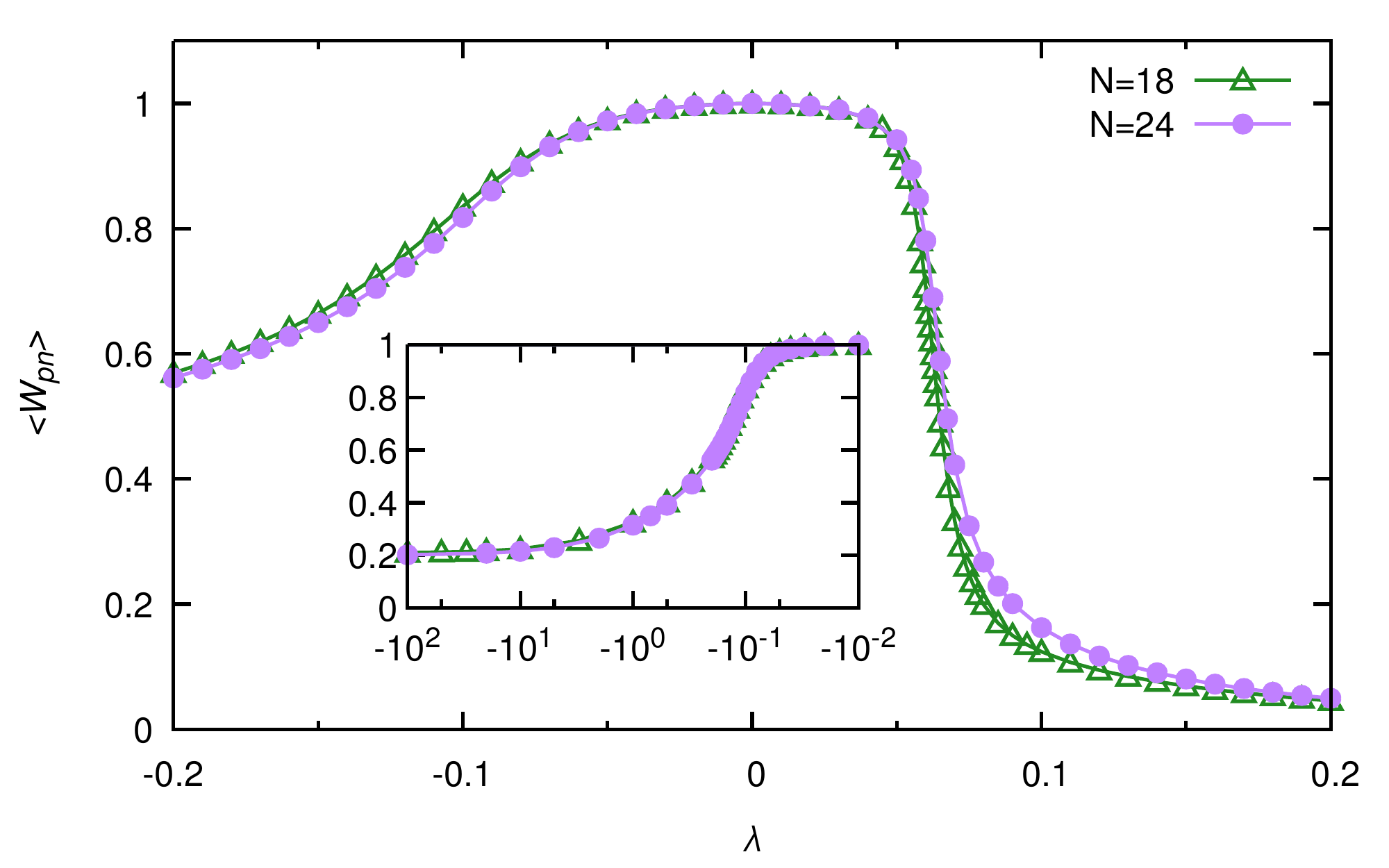}
\caption{
Expectation value of $W_{pn}$ as a function of $\lambda$
  in the bilayer Kitaev model.
}
\label{fig:wp}
\end{figure}

Figure~\ref{fig:wp} shows the expectation value of $\langle W_{pn}\rangle$.
%%%%%%%%%%%%%%%%%%%%%%%%%%%%%%%%%%%%
%%%%%%%%%%%%%%%%%%%%%%%%%%%%%%%%%%%%
This is a conserved quantity in the monolayer Kitaev model and thereby
$W_{pn}=1$ for $\lambda=0$.
In the bilayer Kitaev model, the quantity is no longer conserved,
but it should be appropriate
to discuss how the QSL state realized in monolayer Kitaev model survives
by the introduction of the interlayer coupling.
We find the rapid change around $\lambda\sim 0.06$, suggesting
the existence of the quantum phase transition, as discussed above.
On the other hand, in the negative $\lambda$ region,
this quantity smoothly changes.
In the $\lambda\rightarrow -\infty$ limit,
$\langle W_{pn}\rangle\sim 0.2$.
This implies that the QSL state in the monolayer Kitaev model
is adiabatically connected to the QSL state in the $S=1$ Kitaev model.

%\subsection{dynamical spin structure factor}
%%%%%%%%%%%%%%%%%%%%%%%%%%%%%%%%%%%%
\begin{figure}[htb]
\centering
\includegraphics[width=\columnwidth]{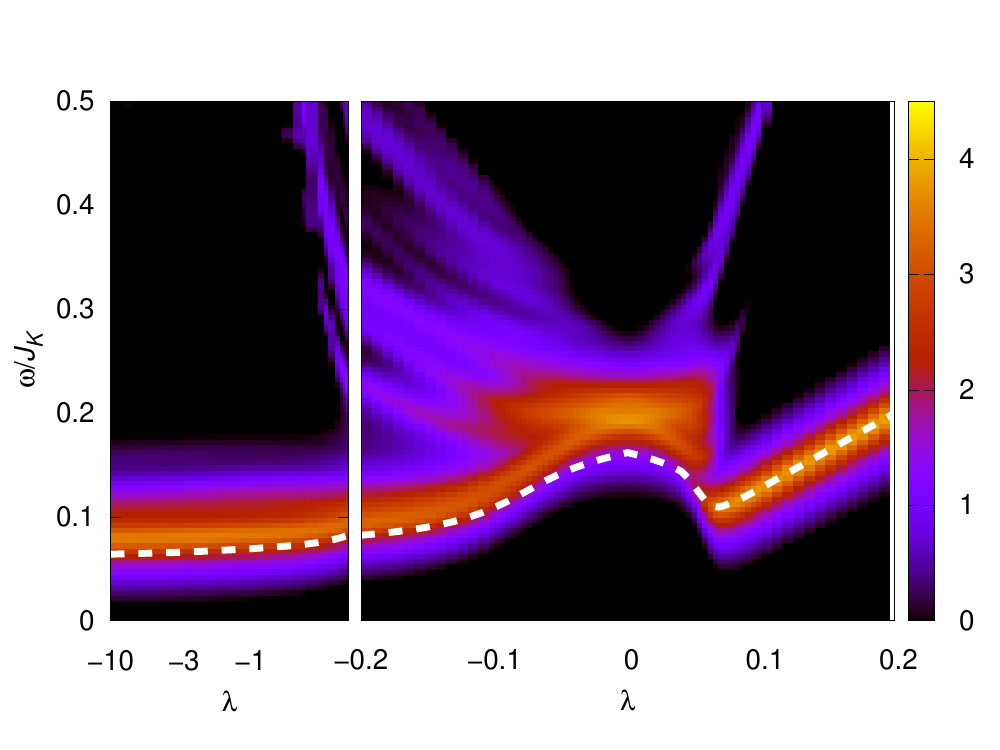}
\caption{
Dynamical spin structure factor $\ln S_n({\bf q}=0,\omega)$
for the $N=18$ system with $\delta^+/J_K=10^{-2}$.
The dashed line is the energy gap $\Delta = E[{\cal S}_1]-E_0$ (see text).
}
\label{fig:sqw}
\end{figure}
%%%%%%%%%%%%%%%%%%%%%%%%%%%%%%%%%%%%
To study ground state properties in more detail, we focus on spin excitations in the system.
%Since excitation structures are well studied for the monolayer Kitaev models,
%we can compare them with those of the present model.
%To clarify spin excitations,
We calculate the dynamical spin structure factor $S({\bf q}=0,\omega)=\sum_\alpha S_n^{\alpha\alpha}({\bf q}=0,\omega)$
for the bilayer Kitaev model with $N=16$ and $18$.
Since their qualitative difference could not be found,
we show in Fig.~\ref{fig:sqw} the dynamical structure factor
for the $N=18$ system.
%The gap behavior appears
It is found that it takes a large value in the low energy region.
Namely, the gap magnitude for the monolayer model $\lambda=0$ is
twice larger than the exact result~\cite{Knolle2014}
due to the finite size effect.
%although the gap magnitude was overestimated due to the finite size effect.
%
An important point is that the sudden change in the spin structure factor appears
around $\lambda\sim 0.05$.
This strongly suggests the existence of the quantum phase transition.
Since the spin gap is proportional to the exchange coupling $J_H$
in the dimer state $(\lambda>0.05)$, linear behavior appears in the peak position of the quantity.
On the other hand, no clear singularity appears
in the lowest energy excitations
in the negative $\lambda$ region, in contrast to the positive case.
This means that the crossover occurs between two QSL states
at $\lambda=0$ and $\lambda=-\infty$.
%When the ferromagnetic interlayer interaction increases,
%the dynamical structure factor has the low-energy peak around $\omega\sim 0.07 J_K=0.14J_{\rm eff}$.
  To understand this spin excitation in the negative $\lambda$ region,
  we also evaluate the ground-state energy in the subspace ${\cal S}_1$
  where %the conserved quantities non-commuting with a local spin are
  $X_p=-1$ for the adjacent two plaquttes $p$
  and $X_{p'}=+1$ for the other plaquettes $p'$.
  The energy gap $\Delta = E[{\cal S}_1]-E_0$ is shown as the dashed line in Fig.~\ref{fig:sqw}.
  We find that the obtained energy gap is consistent with the lower edge of
  the low-energy peak in the dynamical structure factor.
  Note that this behavior is similar to that in the $S=1/2$ case \cite{Knolle2014},
  where the sharp peak at low temperatures appear slightly above the flux gap.
  We have also confirmed that the low-energy spin excitation at $\lambda\to -\infty$
  coincides with that of the $S=1$ Kitaev model in the 18-site cluster.
  The results suggest that the decoupled Kitaev QSL state realized in each layer adiabatically
  connects to the QSL state realized in the $S=1$ Kitaev model.

%%%%%%%%%%%%%%%%%%%%%%%%%%%%%%%%%%%%%%%%%%%%%%
%\subsection{thermodynamic quantities}
%%%%%%%%%%%%%%%%%%%%%%%%%%%%%%%%%%%%%%%%%%%%%%
Next, we discuss thermodynamic properties in the bilayer Kitaev model.
It is known that, in the monolayer Kitaev model,
the double peak structure results from the spin fractionalization in the specific heat.
Similar double peak structure has also been found in the $S=1$ Kitaev model
and the plateau at $S=1/2 \ln 3$ appears in the entropy curve.
Now, we discuss how these double peak structures are connected to each other.
%%%%%%%%%%%%%%%%%%%%%%%%%%%%%%%%%%%%
\begin{figure}[htb]
\centering
\includegraphics[width=\columnwidth]{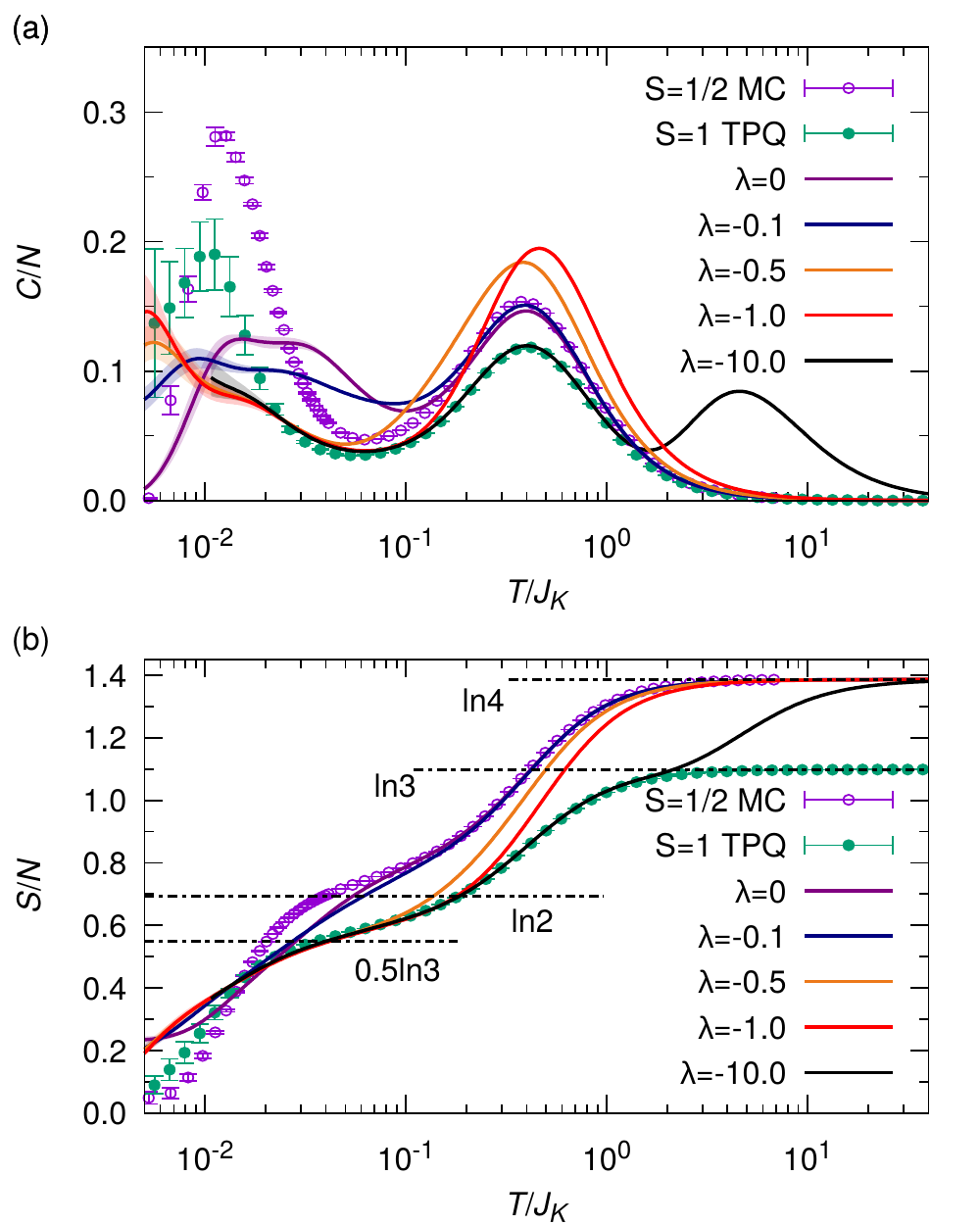}
\caption{
The specific heat per dimer $C/N$, for (a) $\lambda\geq0$ and
(b) $\lambda\leq0$ region, respectively,
as a function of the temperature for the $N=16$ cluster.
Shaded areas are the possible errors estimated by
the standard deviation of the results calculated by 10 TPQ states.
The data for $S=1/2$ and $S=1$ are obtained from the Monte Carlo simulations with $N=800$  sites~\cite{PhysRevB.92.115122,PhysRevLett.119.127204} and the TPQ state method with $N=18$ sites~\cite{koga2018ground}.
}
\label{fig:tpq}
\end{figure}
%%%%%%%%%%%%%%%%%%%%%%%%%%%%%%%%%%%%
By using the TPQ state method starting from, at least, ten initial vectors,
we deduce the thermodynamic quantities at finite temperatures.
Figure~\ref{fig:tpq} shows the temperature dependent specific heat and entropy per dimer.
When $\lambda=0$,
the system is reduced to the decoupled monolayer Kitaev models.
In the case, the double-peak structure appears
around $T/J_K\sim 0.02$ and $0.5$ in the specific heat and
the plateau structure at $S=\ln 2$ in the entropy curve.
This is consistent with the Monte Carlo data~\cite{PhysRevB.92.115122,PhysRevLett.119.127204}
although finite size effects appears at low temperatures $(T/J_K<0.1)$.
As the interlayer interaction is introduced,
the specific heat increases at intermediate temperatures $0.1\lesssim T/J_K\lesssim 0.5$.
This implies that the plateau region at $S/N=\ln 2$ smears in the entropy curve.
In fact, we could not find such a plateau when $J_H/J_K\lesssim -0.5$.
On the other hand, the double peak structure in the specific heat still appears
even when $\lambda\lesssim -0.5$.
This yields another plateau around $T/J_K\sim 0.08$ in the entropy curve.
An important point is that, at low temperatures $T/J_K\lesssim 0.08$,
the entropy data with $\lambda<-0.5$ are almost identical.
This means that low energy magnetic properties should be described by
the $S=1$ Kitaev model Eq.~(\ref{eq:eff}).
Further decreasing $\lambda$ induces another plateau in the entropy $S/N\sim \ln 3$.
When $\lambda=-10$,
this plateau is clearly seen around $T/J_K\sim0.2$, and
the corresponding peak in the specific heat
appears at a relatively higher temperature $T/J_K\sim 5$.
This characteristic temperature should be scaled by the exchange coupling $J_H$.
Therefore, this peak indicates the formation of the spin triplet states,
which is consistent with the fact that
the corresponding entropy $\ln 4-\ln 3$ is released with decreasing temperatures.

%%%%%%%%%%%%%%%%%%%%
\section{Summary}
%%%%%%%%%%%%%%%%%%%%
In summary, we have investigated the ground-state and finite-temperature properties of the bilayer Kitaev model
using the exact diagonalization up to the 24 dimer sites.
The obtained results for the ground-state properties---the ground-state energy,
local quantity defined on each plaquette, and dynamical spin structure factor---suggest that a quantum phase transition occurs
between the Kitaev QSL and dimer singlet states
when the interlayer coupling is antiferromagnetic.
We would like to note that this is in agreement with our previous paper.
On the other hand, the results for the ferromagnetic interlayer coupling indicate
that the $S=1/2$ and $S=1$ Kitaev QSL states adiabatically connect to one another.
In addition, we have discussed the thermodynamic quantities
using the TPQ state method.
The double-peak structure in the specific heat,
which is intrinsic to the $S=1/2$ Kitaev QSL state,
is confirmed to be sustained in the crossover region between the two types of Kitaev QSL states.
The present results are expected to stimulate studies on generalized spin-$S$ Kitaev physics.

\begin{acknowledgments}
%We thank ***.
Parts of the numerical calculations were performed
in the supercomputing systems in ISSP, the University of Tokyo.
This work was supported by Grant-in-Aid for Scientific Research from
JSPS, KAKENHI Grant Nos. JP18K04678, JP17K05536 (A.K.),
JP16K17747, JP16H02206, JP18H04223 (J.N.).
\end{acknowledgments}

%%%%%%%%%%%%%%%%%%%%%%
\bibliography{./refs}
%%%%%%%%%%%%%%%%%%%%%%
\end{document}